# Magnetic-field-induced quantized anomalous Hall effect in intrinsic magnetic topological insulator MnBi$_2$Te$_4$

Yujun Deng[1,3,†], Yijun Yu[1,3,†], Meng Zhu Shi[2,4,†], Jing Wang[1,3*], Xian Hui Chen[2,4*] and Yuanbo Zhang[1,3*]

[1]*State Key Laboratory of Surface Physics and Department of Physics, Fudan University, Shanghai 200438, China*

[2]*Hefei National Laboratory for Physical Science at Microscale and Department of Physics, University of Science and Technology of China, and Key Laboratory of Strongly-coupled Quantum Matter Physics, Chinese Academy of Sciences, Hefei, Anhui 230026, China*

[3]*Institute for Nanoelectronic Devices and Quantum Computing, Fudan University, Shanghai 200433, China*

[4]*Key Laboratory of Strongly Coupled Quantum Matter Physics, University of Science and Technology of China, Hefei, Anhui 230026, China*

[†] These authors contributed equally to this work.

[*] Correspondence should be addressed to Y.Z. (zhyb@fudan.edu.cn), X.H.C. (chenxh@ustc.edu.cn) and J.W. (wjingphys@fudan.edu.cn).

**In a magnetic topological insulator, nontrivial band topology conspires with magnetic order to produce exotic states of matter that are best exemplified by quantum anomalous Hall (QAH) insulators (*1–5*) and axion insulators (*2*, *6–8*). Up till now, such magnetic topological insulators are obtained by doping topological insulators with magnetic atoms (*9*). The random magnetic dopants, however, inevitably introduce disorders that hinder further exploration of topological quantum effects in the material. Here, we resolve this dilemma by probing quantum transport in MnBi$_2$Te$_4$ thin flake—a topological insulator with intrinsic magnetic order. In this layered van der Waals crystal, the ferromagnetic layers couple anti-parallel to each other, so MnBi$_2$Te$_4$ is an antiferromagnet (*10–13*). A magnetic field, however, aligns all the layers and induces an interlayer ferromagnetic order; we show that a quantized anomalous Hall response emerges in atomically thin MnBi$_2$Te$_4$ under a moderate magnetic field. MnBi$_2$Te$_4$ therefore becomes the first intrinsic magnetic topological insulator exhibiting quantized anomalous Hall effect. The result establishes MnBi$_2$Te$_4$ as an ideal arena for further exploring various topological phenomena.**

The discovery of topological quantum materials demonstrates the importance of band topology that is increasingly recognized in condensed matter physics (*9*, *14*, *15*). A distinct feature common to all topological materials is the presence of topologically protected quantum states that are robust against local perturbations. For example, in a topological insulator (TI) such as Bi$_2$Te$_3$, the bulk band topology guarantees the existence of gapless two-dimensional (2D) surface states with gapless Dirac dispersion (*16*, *17*). Introducing magnetism into the initially time-reversal invariant TIs brings about profound changes in their electronic structure. Specifically, the long-range magnetic order breaks the time-reversal symmetry, and causes an exchange gap in the gapless Dirac dispersion of the surface states (*2*, *18*). The gap opening is accompanied by the emergence of a chiral edge mode that was predicted to give rise to a QAH effect when the Fermi level is situated in the exchange gap (*2*). The dissipationless QAH edge channel, combined with the spin-momentum locking inherent in topological materials, may lead to new device concepts for topological electronic applications (*9*).

The experimental observation of the QAH effect in chromium-doped $(Bi,Sb)_2Te_3$ (*5, 19–21*) represents a triumph in topological quantum material research—the ratio of the multiple elements in the non-stoichiometric material has to be precisely controlled to accomplish such a feat. The fine-tuning required to reconcile conflicting demands, i.e. large magnetization and low initial carrier doping, poses a great challenge for material growth. To make matters worse, the randomly distributed magnetic dopants used to achieve ferromagnetism also act as impurities that limit the quality of the magnetic TIs. As a result, the exact quantization of the anomalous Hall effect appears only at low temperatures; the current record is limited to 2 K (in penta-layer sandwich structure of topological insulators) (*22*), far below the ferromagnetic Curie temperature (a few tens of Kelvin) and exchange gap ($\sim 30 \text{ meV}$) (*23*) in the material. Further exploration of rich topological phenomena and their potential applications calls for intrinsic magnetic TIs—stoichiometric TIs with an innate magnetic order such that topological effects can be studied in pristine crystals.

Here we probe the quantum transport in atomically thin flakes of intrinsic magnetic TI $MnBi_2Te_4$. $MnBi_2Te_4$ is layered ternary tetradymite compound that consists of Te-Bi-Te-Mn-Te-Bi-Te septuple layers (SLs), so the material can be viewed as layered TI $Bi_2Te_3$ with each of its Te-Bi-Te-Bi-Te quintuple layer intercalated by an additional Mn-Te bilayer (Fig. 1A). The resultant $MnBi_2Te_4$ crystal remains a TI, but now becomes intrinsically magnetic (*10–13*). The magnetism originates from the $Mn^{2+}$ ions in the crystal, which have a high spin of $S = 5/2$ and a large magnetic moment of $\sim 5\mu_B$ ($\mu_B$ is the Bohr magneton) (*10, 12*). Below a Neel temperature of $T_N = 25$ K, the spins couple ferromagnetically in each SL with an out-of-plane easy axis, but adjacent SLs couple anti-parallel to each other; bulk $MnBi_2Te_4$ is therefore an antiferromagnet (AFM) (*10, 12*). In this work, we focus on thin flakes of $MnBi_2Te_4$ to minimize the parallel bulk conduction. Because the SLs are separated by van der Waals gaps in bulk $MnBi_2Te_4$, the extensive arsenal of fabrication techniques developed for 2D materials enable us to obtain few-layer samples that preserve the high quality of the crystals. As we show below, a moderate perpendicular magnetic field is able to sequentially flip the ferromagnetic SLs, and eventually drive the material into a ferromagnetically ordered state; a well quantized anomalous Hall effect is realized at a temperature as high as 4.5 K.

We start with high quality $MnBi_2Te_4$ single crystals that are grown by self-flux method (*12, 24*), and obtain atomically thin $MnBi_2Te_4$ using an $Al_2O_3$-assisted exfoliation technique described in (*25*). Specifically, we first thermally evaporate $Al_2O_3$ thin film onto the freshly-prepared surface of bulk crystal. We then lift the Al$_2$O$_3$ thin film, along with MnBi$_2$Te$_4$ thin flakes cleaved from the bulk, using a thermal release tape. The Al$_2$O$_3$/MnBi$_2$Te$_4$ stack is subsequently released onto a piece of transparent polydimethylsiloxane (PDMS), and inspected under optical microscope in transmission mode. Fig. 1B displays the optical image of few-layer MnBi$_2$Te$_4$ flakes on the Al$_2$O$_3$ film attached to PDMS. The transmittance of various number of SLs follows the Beer-Lambert law (Fig. 1C), which enables us to precisely determine the layer number. The thin flakes are finally stamped onto Si wafer covered with 285 nm-thick SiO$_2$, followed by deposition of metal contacts for transport measurements. The degenerately doped Si serves as a back gate, so a voltage bias $V_g$ applied between Si and the sample can electrostatically dope electron or hole carriers into MnBi$_2$Te$_4$ flakes, depending on the polarity of $V_g$. The entire device fabrication process is performed in an argon-filled glove box where O$_2$ and H$_2$O content is kept below 0.5 parts per million to mitigate sample degradation. We focus on flakes with odd number of SLs, which are intrinsically ferromagnetic due to unbalanced layer polarization.

Figure 1D displays temperature-dependent resistance, $R_{xx}$, of few-layer MnBi$_2$Te$_4$. A prominent feature in the data sets is the resistance peak (or kink for the seven-layer sample) at low temperature. Similar peak was also observed in temperature-dependent resistance of the bulk crystal, where it is attributed to increased spin-scattering at the AFM transition (*26*). The location of the peak, therefore, gives a measure of the Neel temperature $T_N$ in the few-layer specimens. We find that the seven-layer sample has a $T_N$ of 24 K, comparable to the value in the bulk. The $T_N$ is however slightly suppressed in thinner samples (23 K for the five-layer sample and 18 K for the three-layer sample), which can be ascribed to increased thermal fluctuations as the samples approach the 2D limit. The resistances rise again at lowest temperatures, probably because of localization of the carriers in the presence of defects in the samples.

Hall measurements provide further information on the quality of the thin flakes (Fig. 1E). First of all, the slope of the Hall resistance $R_{yx}$ near zero magnetic field, referred to as Hall coefficient $R_H$, yields the initial carrier concentration in the sample, $n = 1/eR_H$ ($e$ is the charge of an electron). We find that all samples are initially electron doped; the five-layer and three-layer samples have initial dopings of $7 \times 10^{11}$ cm$^{-2}$ and $2 \times 10^{12}$ cm$^{-2}$, respectively, whereas the initial doping level of the seven-layer sample is more than one order of magnitude higher. Such electron doping is probably induced by antisite defects and/or Mn vacancies in the crystal (*27*), and the doping level gives a qualitative measure of the defect density. In the following experiments, we mainly focus on the five-layer and three-layer samples that have low carrier concentrations and high quality. Second, the Hall mobility $\mu_\text{H}$ of the samples can be determined from $\mu_\text{H} = \sigma/ne$, where $\sigma = R_{xx}W/L$; $W$ and $L$ are the width and length of the sample, respectively. All our samples exhibit mobility values ranging from 100 - 1000 cm$^2$ V$^{-1}$ s$^{-1}$, on the same order as the magnetic TI thin films grown by molecular beam epitaxy (*5*, *19*, *28*). The high quality of the samples is also reflected in the sharp ferromagnetic transition, which implies that the samples (with a typical size of $10 \times 10$ μm$^2$) flip as a single domain at the coercive field. Finally, we note that the zero-magnetic-field anomalous Hall response in the five-layer and three-layer samples became a significant fraction of the quantum resistance $h/e^2$ ($h$ is the plank constant), but did not quantize; a larger exchange gap is needed to overcome disorder and temperature fluctuations.

An external magnetic field applied perpendicular to the sample aligns the magnetization in all SLs to induce a large exchange gap (Figs. 2C and 2D). We observe the exact quantization of anomalous Hall effect in a pristine five-layer MnBi$_2$Te$_4$ flake under a moderate magnetic field. Figs. 2A and 2B display $R_{xx}$ and $R_{yx}$ of the sample as a function of magnetic field recorded at various temperatures. At the lowest temperature of $T = 1.6$ K, $R_{yx}(\mu_0 H)$ reaches a quantized plateau at $h/e^2$ above $\mu_0 H \sim 6$ T, while $R_{xx}$ approaches zero at the same time; both features are hallmarks of quantized Hall effect that signifies the emergence of a chiral 1D dissipationless state at the edges of the sample (Fig. 2D). The magnetic-field-induced quantized Hall effect in five-layer MnBi$_2$Te$_4$ is robust at elevated temperatures. We observe that $|R_{yx}|$ stays above $0.97h/e^2$ (and $R_{xx}$ remains below $0.017h/e^2$) under a magnetic field of $\mu_0 H = 12$ T at temperatures up to $T = 4.5$ K (Fig. 2E). This quantization temperature is significantly higher than the highest value found in modulation-doped magnetic TI thin films (*22*, *29*). At higher temperatures, $R_{yx}$ deviates from exact quantization, and meanwhile $R_{xx}$ exhibits a thermally activated behavior, $R_{xx} \propto \exp(-\Delta/2k_\text{B}T)$ ($k_\text{B}$ is the Boltzmann constant and $\Delta$ is the activation gap; Fig. 2E). Line fit to the Arrhenius plot of $\ln R_{xx}$ as a function of $1/T$ yields $\Delta = 21$ K (Fig. 2E, black line). Even though this energy scale is again much larger than that in magnetic TI thin films (*21*), we note that $\Delta$, which measures the energy gap between the mobility edges, is still much lower than the exchange gap (up to $\sim 88$ meV) in MnBi$_2$Te$_4$ (*11*, *12*, *24*, *26*). There is therefore much room for further increasing $\Delta$ in pure crystals of MnBi$_2$Te$_4$.

It may be asked that why the quantized Hall effect observed here is not caused by Landau level quantization as in the case of the ordinary quantum Hall effect—after all, a fully-filled first Landau level also produces exact quantization of $|R_{yx}| = h/e^2$. Close examination on the origin of the quantization, however, reveals fundamental difference between the two variants of quantized Hall effects that can be probed experimentally. In both cases, the Hall conductivity $\sigma_{xy}$ is determined by a topological invariant integer $C$ known as Chern number, $\sigma_{xy} = Ce^2/h$. For the ordinary quantum Hall effect, $C$ corresponds to the occupancy of the Landau levels, so the sign of $C$ is uniquely determined by the sign of the charge carriers (electrons or holes). In contrast, the sign of $C$ depends only on the sign of the exchange coupling and magnetization direction, but not on the charge carriers, in a QAH insulator (*2*, *9*). Probing the quantized Hall response while switching the sign of charge carriers—but leaving the exchange coupling and magnetization intact—is a good way to unambiguously distinguish the two cases. To this end, we sweep the magnetic field, and record $R_{yx}(\mu_0 H)$ and $R_{xx}(\mu_0 H)$ of a five-layer sample under various back gate biases $V_g$ (Fig. 3A and 3B). As $V_g$ is tuned from zero to $-90$ V, the initially electron-doped sample becomes hole doped, as judged from the sign of the Hall coefficient $R_H$ (Fig. 3C). (The charge neutral point (CNP) is identified at $V_g^{CNP} \approx -50$ V, where $R_H$ switches sign, and zero-field $R_{xx}$ exhibits a peak as shown in Fig. 3D.) We observe that the quantized $R_{yx}$ does not change sign on either side of the CNP, and therefore rule out Landau level quantization as a probable cause of the quantized Hall effect in MnBi$_2$Te$_4$. We note that the $R_{yx}$ quantization becomes better as $V_g$ is tuned close to $V_g^{CNP}$ (Fig. 3A). This observation further indicates that $R_{yx}$ quantization is linked to the chiral edge state located inside of the exchange gap.

The gate-dependent $R_{xx}(\mu_0 H)$ and $R_{yx}(\mu_0 H)$ contains important information on the evolution of the magnetic states in the five-layer MnBi$_2$Te$_4$ flake. There are two main points to notice. First, the peak in the magnetoresistance $R_{xx}(\mu_0 H)$ at the coercive field transforms into a dip as the charge carriers in the sample switch from electrons to holes. The peak on the electron side can be understood as a result of enhanced electron-magnon scattering when the external field is antiparallel to magnetization (*30*). This theory, however, does not explain the dip observed on the hole side. Here we note that the dip feature resembles the butterfly-shaped magneto-resistance observed in another magnetic TI Mn-Bi$_2$Te$_3$ at the dimensional cross-over regime (*31*), where magnetic Skyrmions may form. The same mechanism may be at work in MnBi$_2$Te$_4$ thin flakes.

Second, both $R_{xx}(\mu_0 H)$ and $R_{yx}(\mu_0 H)$ undergo a sequence of transitions as the magnetic field increases, before the sample settles into the quantized anomalous Hall state. The transitions are manifestations of the complex intermediate magnetic states that are precursors of the QAH insulator at high magnetic fields.

Important insight into these magnetic states is gained when we juxtapose the magneto-transport in the five-layer sample with that in a three-layer sample (Fig. 4). The $R_{yx}(\mu_0 H)$ in the three-layer sample exhibits two plateaus that indicate two magnetic states, in contrast with three magnetic states in the five-layer sample. It now becomes clear that the magnetic states are in fact initially antiferromagnetically coupled SLs with individual SLs flipped, one SL at a time, by an increasing external magnetic field. These magnetic states are schematically illustrated in Fig. 4. The transitions between the states can be described by the Stoner-Wohlfarth model with bipartite AFM (*24*). The model further points out that each layer-flip takes place in two steps via a spin flop transition, where the layer magnetization is free to rotate in directions approximately perpendicular to the easy axis (*24*). Close examination of the transitions in $R_{yx}(\mu_0 H)$ indeed reveal signs of such two-step processes.

In summary, we observe a quantized anomalous Hall effect in atomically thin flakes of intrinsic magnetic TI—MnBi$_2$Te$_4$. The quantized anomalous Hall effect is induced by an external magnetic field that drives the initially antiferromagnetically coupled layers to align ferromagnetically; dissipationless 1D chiral edge conduction emerges in the QAH insulator. The quantization is robust at $T = 4.5$ K, and the transport gap is $\Delta = 21$ K. Both values are expected to improve in pure MnBi$_2$Te$_4$ crystals. Because MnBi$_2$Te$_4$ is also a 2D material, the techniques developed for other 2D materials can be readily applied to MnBi$_2$Te$_4$. We anticipate that van der Waals heterostructures integrating MnBi$_2$Te$_4$ with other magnetic/superconducting 2D materials will provide fertile ground for exploring exotic topological quantum phenomena.

**Acknowledgements**

We thank Xiaofeng Jin and Yizheng Wu for helpful discussions. Part of the sample fabrication was conducted at Nano-fabrication Laboratory at Fudan University. Y.D., Y.Y., J.W. and Y.Z. acknowledge support from National Key Research Program of China (grant nos. 2016YFA0300703, 2018YFA0305600), NSF of China (grant nos. U1732274, 11527805, 11425415 and 11421404), Shanghai Municipal Science and Technology Commission (grant no. 18JC1410300), and Strategic Priority Research Program of Chinese Academy of Sciences (grant no. XDB30000000). Y.Y. also acknowledges support from China Postdoctoral Science Foundation (grant no. BX20180076 and 2018M641907). J.W. also acknowledges support from the NSF of China (grant no. 11774065) and NSF of Shanghai (grant no. 17ZR1442500). M.Z.S. and X.H.C. acknowledge support from the National Natural Science Foundation of China (grant nos. 11888101, 11534010), the National Key R&D Program of China (grant nos. 2017YFA0303001 and 2016YFA0300201), Strategic Priority Research Program of the Chinese Academy of Sciences (grant no. XDB25000000), and the Key Research Program of Frontier Sciences, CAS (grant no. QYZDY-SSW-SLH021).

**Author Contributions**

Y.Z., X. H. C. and J. W. supervised the project. M.Z.S. and X.H.C. grew the MnBi$_2$Te$_4$ bulk crystals. Y.D. and Y.Y. fabricated devices and performed measurements. Y.D., Y.Y., Y.Z., J.W. and X. H. C. analyzed the data. J.W. carried out theoretical calculations. Y.D., Y.Y., J.W. and Y.Z. wrote the manuscript with input from all authors.

**Figure captions**

**Fig. 1. Fabrication and characterization of few-layer MnBi₂Te₄ devices.** (**A**) Crystal structure of MnBi₂Te₄. The septuple atomic layers are stacked together by van der Waals interactions. Red and blue arrows denote magnetic moment of each Mn$^{2+}$ ion. Under zero external magnetic field, the neighboring SLs are antiferromagnetically coupled with an out-of-plane magnetocrystalline anisotropy. (**B**) Optical image of few-layer flakes of MnBi₂Te₄ cleaved by thermally-evaporated Al₂O₃ thin film. The MnBi₂Te₄/Al₂O₃ stack is supported on a piece of PDMS. Number of SLs are labelled on selected flakes. Scale bar: 20 μm. (**C**) Transmittance as a function of number of SLs. The data follow the Beer-Lambert law (solid line), which enables us to precisely determine the number of layers. (**D**) Temperature-dependent sample resistance of few-layer MnBi₂Te₄. The antiferromagnetic transition manifests as resistance peaks in the three-layer and five-layer samples (and a kink in the seven-layer sample). (**E**) Hall resistance of few-layer MnBi₂Te₄ under a low external magnetic field. The flakes with odd number of layers are ferromagnetic at low temperatures. Data were obtained at $T = 1.6$ K.

**Fig. 2. Magnetic-field-induced quantized anomalous Hall effect in a five-layer MnBi₂Te₄ sample.** (**A** and **B**) Magnetic-field-dependent $R_{xx}$ and $R_{yx}$ acquired at various temperatures. From bottom: $T = 1.6, 3.0, 5.0, 7.0, 10, 13, 18, 24, 30, 40$ K. Quantized Hall plateaus are observed above $\mu_0 H \sim 6$ T at low temperatures, and the quantization is accompanied by vanishing $R_{xx}$. (**C**) Schematic illustration of the gapped Dirac-like dispersion of the surface state in ferromagnetic MnBi₂Te₄. A chiral edge state (gray line) emerges inside of the exchange gap. Arrows mark the direction of the spins that are locked with crystal momentum. (**D**) Real-space picture of the chiral edge state when an external magnetic field $\mu_0 H$ forces all SLs to align ferromagnetically. (**E**) $R_{yx}$ (right axis) and Arrhenius plot of $\ln R_{xx}$ (left axis) as a function of $1/T$. External magnetic field is fixed at $\mu_0 H = 12$ T. Black solid line is a line fit to the Arrhenius plot, from which we obtain a transport gap of $\Delta = 21$ K.

**Fig. 3. Gate-tuned Magneto-transport a five-layer MnBi₂Te₄ sample.** (**A** and **B**) Magnetic-field-dependent $R_{yx}$ and $R_{xx}$ measured under various gate biases $V_g$. Here $R_{xx}$ ($R_{yx}$) curves

are symmetrized (anti-symmetrized) to separate the two components. (**C**) Hall coefficient $R_H$ plotted as a function of $V_g$. $R_H$ is extracted from fitting $R_{yx}$ near zero magnetic field. As $V_g$ is tuned from zero to $-90$ V, the initially electron doped sample becomes hole doped. $V_g = -50$ V marks the CNP where $R_H$ (and the charge carriers) changes sign. (**D**) Zero-field $R_{xx}$ plotted as a function of $V_g$. $R_{xx}$ exhibit a peak around the CNP. All data were obtained at $T = 1.6$ K.

**Fig. 4. Magnetic transitions in few-layer MnBi₂Te₄ induced by an external magnetic field.** (**A** and **B**) Symmetrized $R_{yx}$ and anti-symmetrized $R_{xx}$ of a three-layer sample (**A**) and a five-layer sample (**B**) shown as a function of external magnetic field applied perpendicularly. All data were obtained at $T = 1.6$ K. The magnetic state at specific locations (marked by black dots) are schematically illustrated. Red marks the spin up SLs and blue marks the spin down SLs. For simplicity, only one of the possible configurations are shown when there is degeneracy.

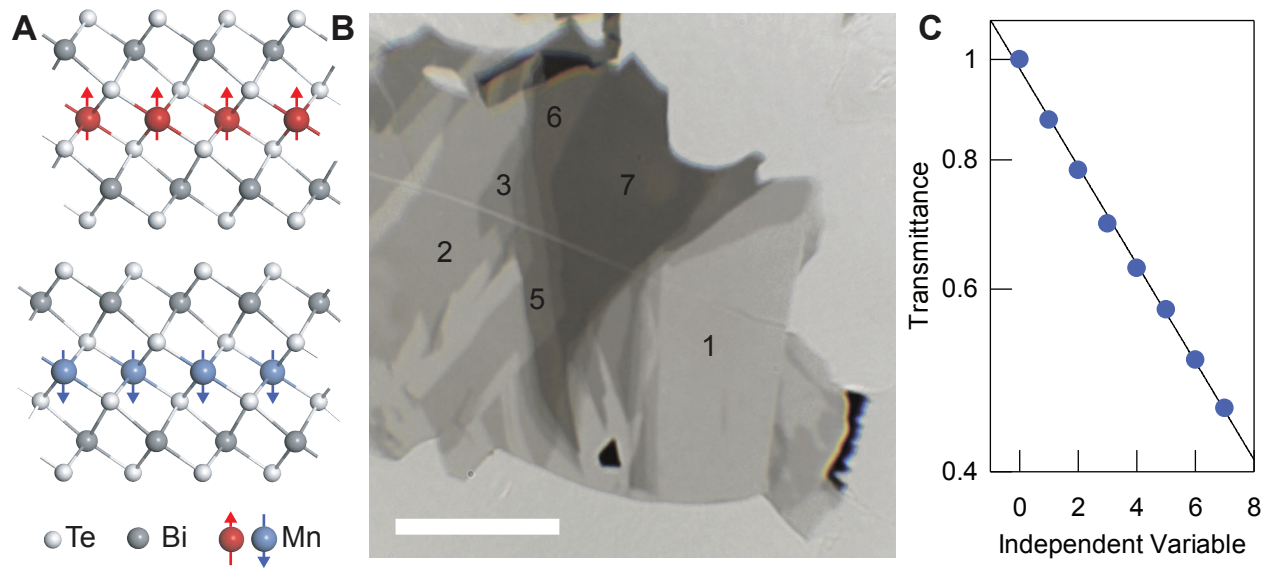
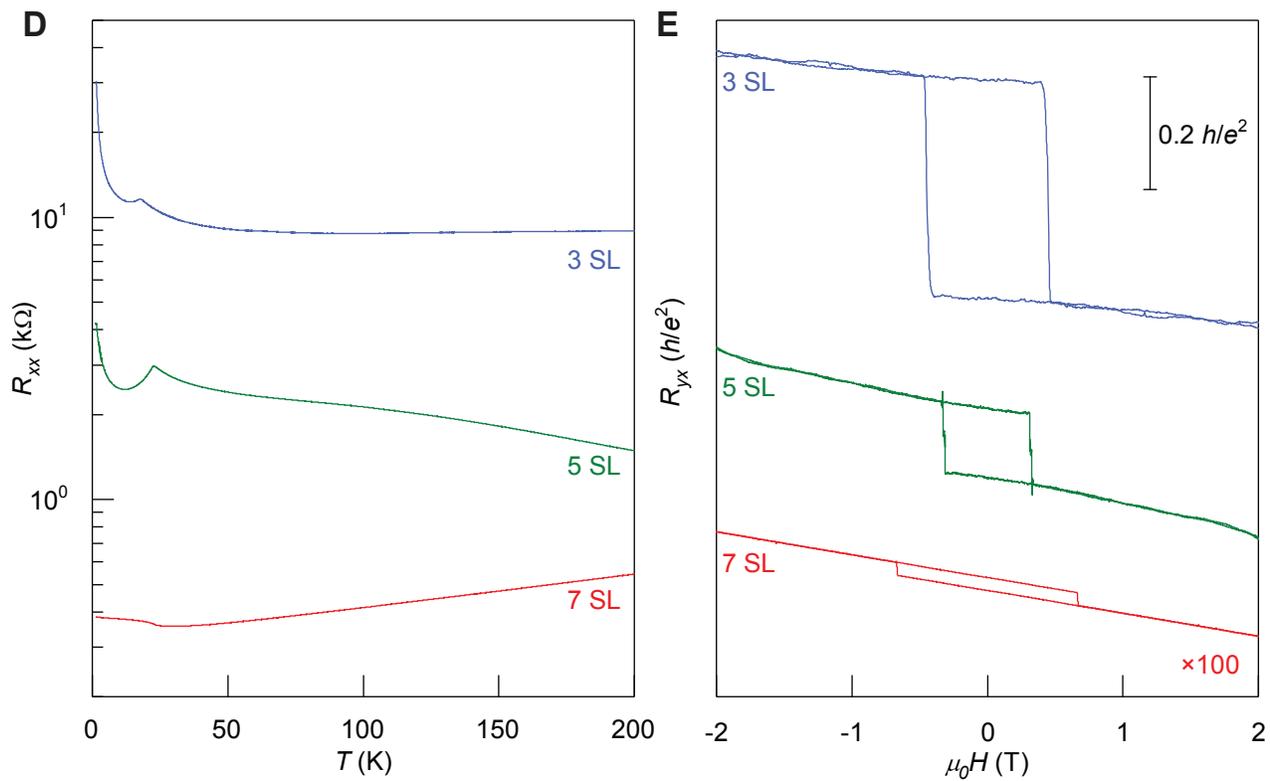

Deng et al., Figure 1

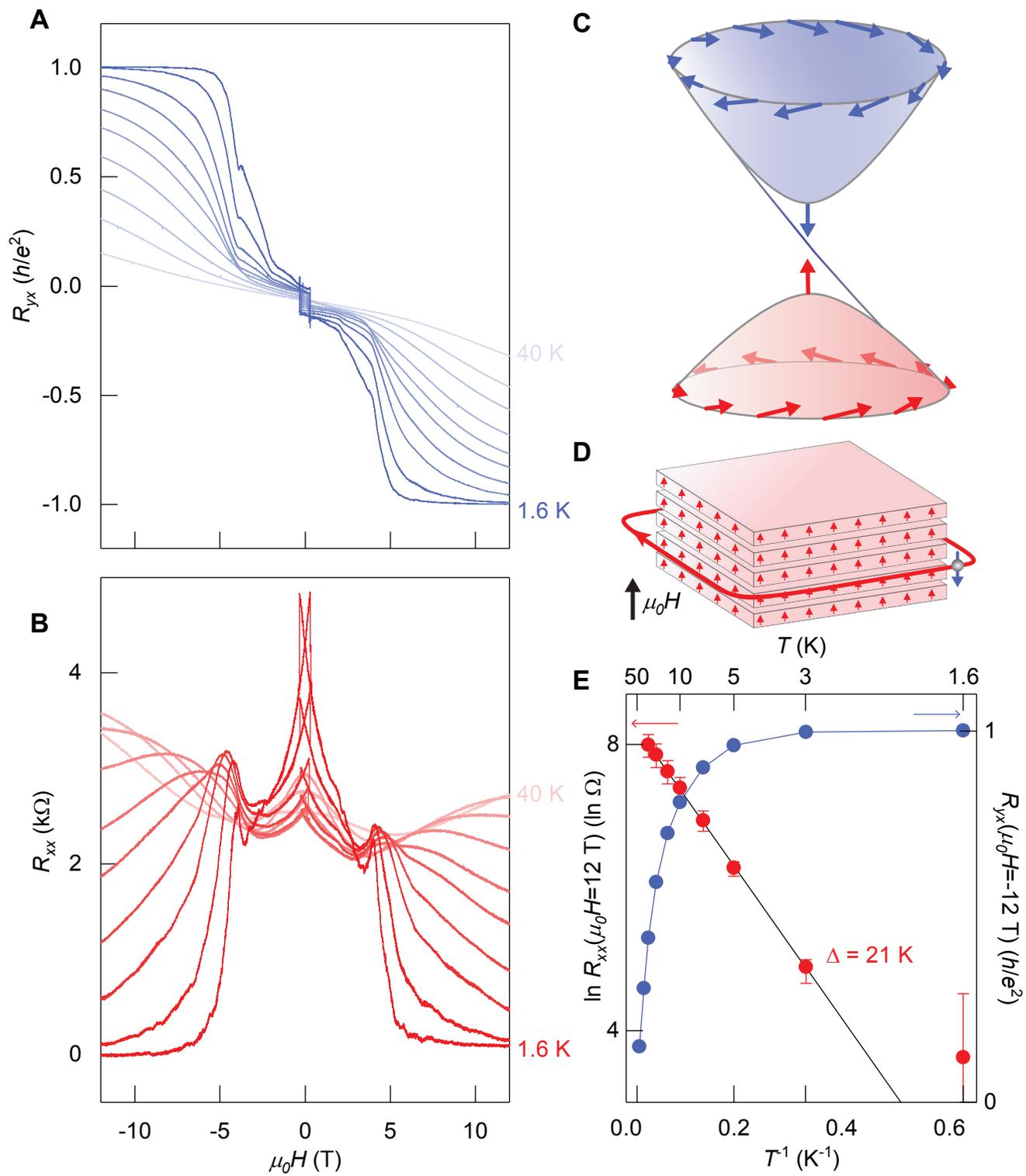

Deng et al., Figure 2

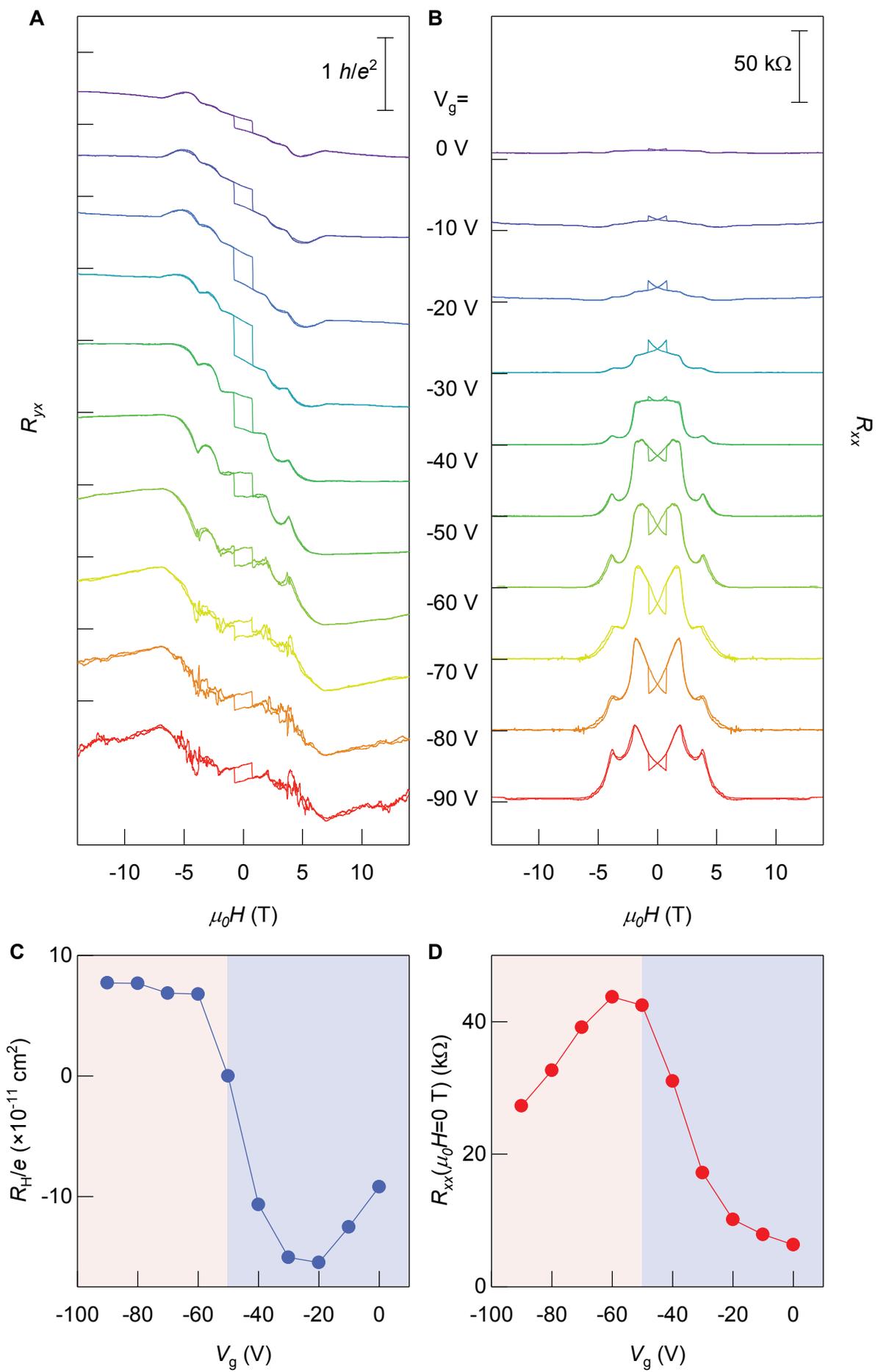

Deng et al., Figure 3

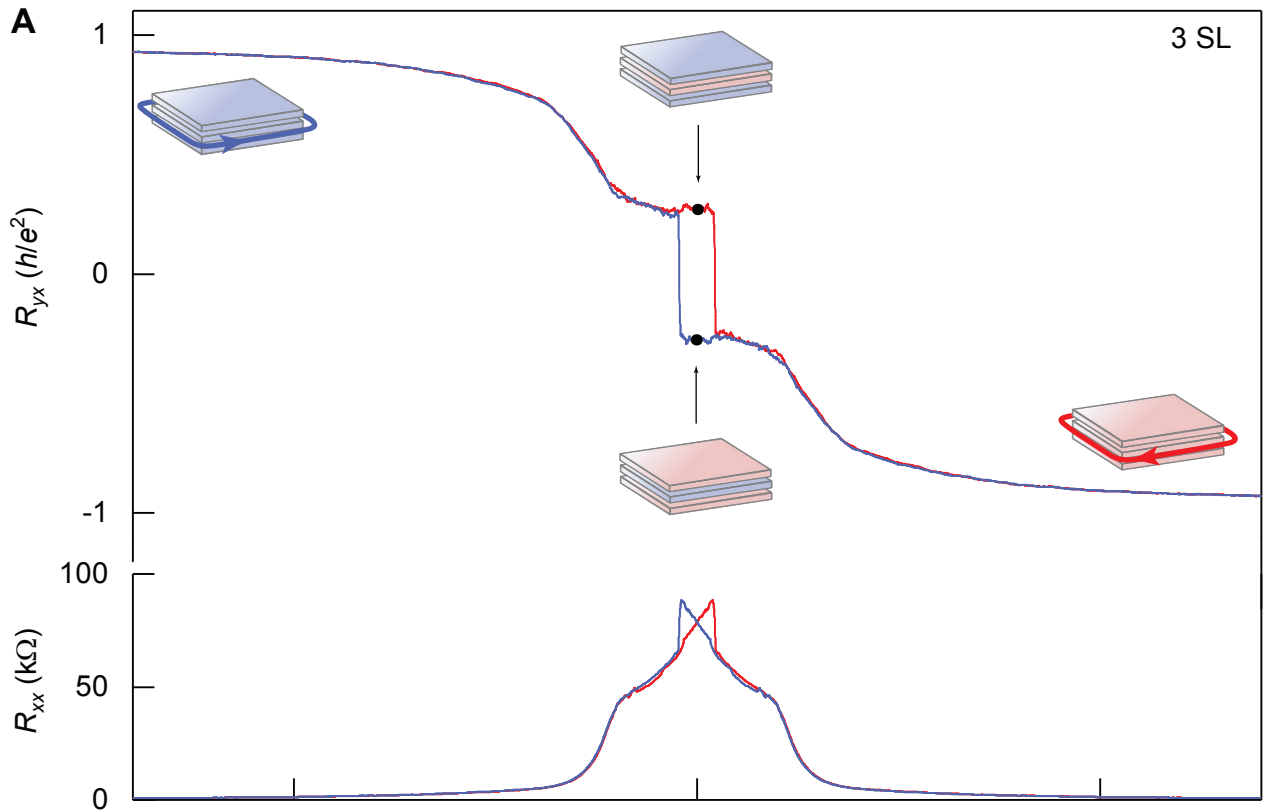
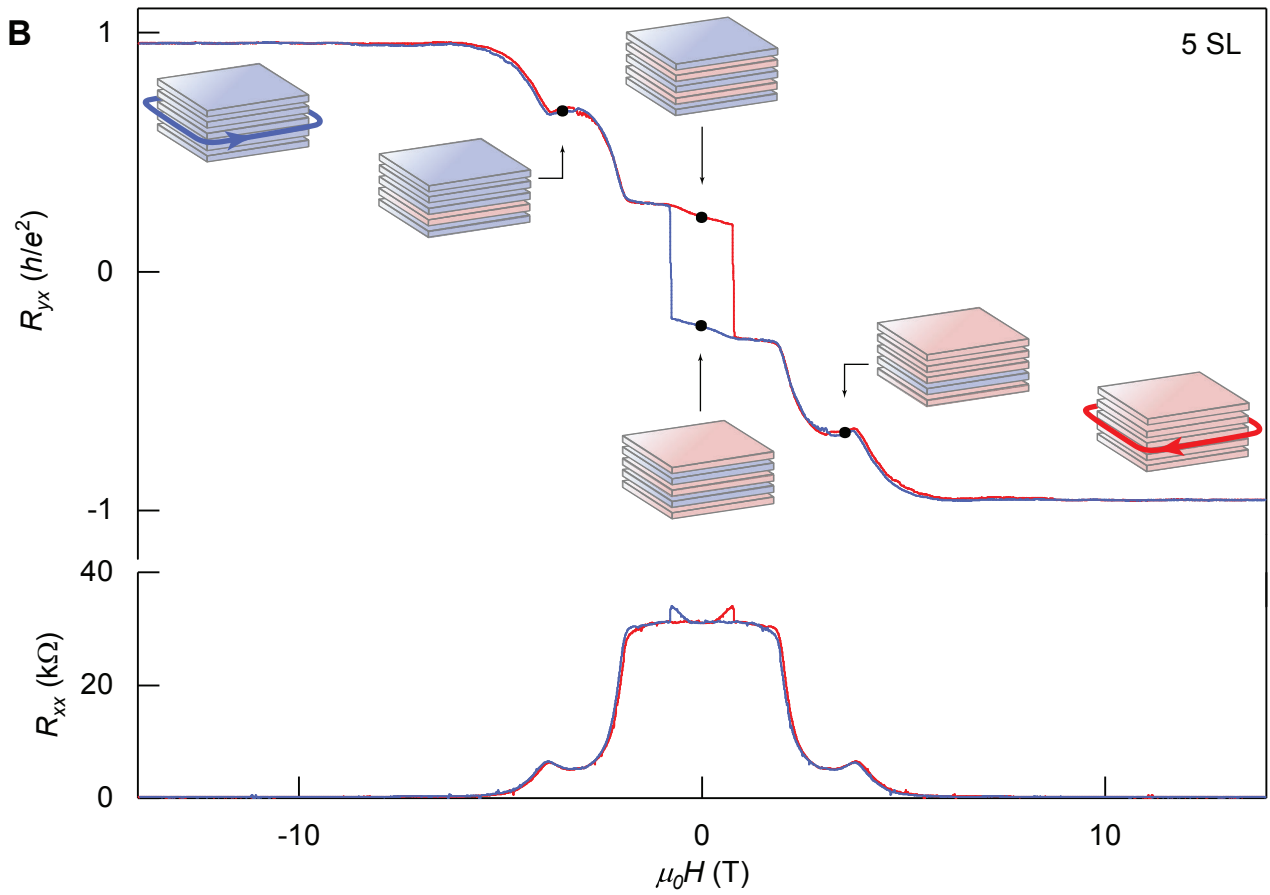

Deng et al., Figure 4